\documentclass{revtex4}

\usepackage{epsfig}

\begin{document}

\title{On the effects of surrogacy of energy dissipation \\
in determining the intermittency exponent in fully developed turbulence}

\author{J.\ Cleve$^{1,2,3}$, M.\ Greiner$^4$, K.\ R.\ Sreenivasan$^{1}$}

\address{$^{1}$Institute for Physical Science and Technology,
        University of Maryland, College Park, MD 20742, USA}
\address{$^2$Max-Planck-Institut f\"ur Physik komplexer Systeme,
        N\"othnitzer Str.\ 38, D--01187 Dresden, Germany}
\address{$^3$Institut f\"ur Theoretische Physik, Technische
        Universit\"at Dresden, D--01062 Dresden, Germany}
\address{$^4$Corporate Technology, Information {\&} Communications,
        Siemens AG, D--81730 M\"unchen, Germany}


\begin{abstract}
The two-point correlation function of the energy dissipation, obtained
from a one-point time record of an atmospheric boundary layer, reveals
a rigorous power-law scaling with intermittency exponent $\mu \approx 0.20$
over almost the entire inertial range of scales. However, for the 
related integral moment, the power-law
scaling is restricted to the upper part of the inertial range
only. This observation is explained in terms of the operational surrogacy of
the construction of energy dissipation, which influences
the behaviour of the correlation function for small separation distances.
\end{abstract}

\maketitle

05.40.-a Fluctuation phenomena, random processes, noise,
                and Brownian motion;\\
02.50.Sk Multivariate analysis\\


Time records of turbulent velocity at a single point in space,
obtained using a hot-wire or a laser Doppler anemometer, are
usually interpreted, via Taylor's frozen flow hypothesis, as
one-dimensional spatial cuts through the flow. Velocity structure
functions can be obtained readily from such observables. In addition
to the velocity, other quantities of interest include enstrophy
and energy dissipation. These quantities cannot be constructed in
full from the measured one-point velocity time series (of one or
two components of velocity) and so are replaced for further
analysis by the so-called
surrogate fields. These surrogate fields
usually take the form of a single component of a many-component
field. In this note we concentrate on the surrogacy issue of
energy dissipation and discuss its impact on the extraction of the
intermittency exponent.

To illustrate the issue, we choose turbulence measurements in an
atmospheric boundary layer, made under nominally steady and nearly
neutral conditions, in which a hot-wire probe mounted on top of a
tower recorded time-series of both streamwise and vertical velocity components;
for details of the experimental setup, see Ref.\ \cite{DHR00}.
The frozen flow hypothesis has been applied to
convert the time series into spatial cuts. Upon using the method
of Ref.\ \cite{LOEF}, the Reynolds number
$R_\lambda = \sqrt{\langle u^2\rangle} \lambda/\nu$, based on the
Taylor
microscale $\lambda = \sqrt{\langle u^2 \rangle/ \langle (\partial
u/\partial x)^2 \rangle}$, was determined to be 9000. The angular
brackets denote
a temporal average throughout the paper. The estimated
ratio between integral length $L$, defined through the integral of
the two-point correlation of the component velocity fluctuation in
the streamwise direction, and the
dissipation scale
$\eta = (\nu^3/\langle\varepsilon\rangle)^{1/4}$, is
$5\times 10^4$. In units of $L$ the record length of the time
series is $L_{\rm record}/L= 1000$. The inertial range of scales
is determined by examining the scaling of the third-order
structure function; within the inertial range so determined, the
power spectra show a
well-defined slope close to $-5/3$. Because of instrument and
cable noise, the
spectral density has some amount of noise contamination
towards the smallest scales. In order to ensure a proper
construction of the derivatives $\partial v_i/\partial x =
(v_i(x+{\Delta}x) - v_i(x))/\Delta x$, the noise part has
been removed from the velocity signal by using a Wiener filter.

The true energy dissipation rate
\begin{equation}
\label{one}
  \varepsilon(\vec{x})
    =  \frac{\nu}{2} \sum_{i,j}(\partial_iv_j+\partial_jv_i)^2,
\end{equation}
where the indices $i$ and $j$ represent the coordinate axes, cannot
be constructed from the recorded time series. This is so because
only
the longitudinal and transverse components $v_x$ and $v_y$ of the
velocity, along the streamwise direction $x$ and the normal direction $y$ respectively, are accessible. Hence,
the true energy dissipation is replaced by a surrogate
proposal. One possibility is
\begin{equation}
\label{two}
  \varepsilon_{\rm surr1}(x)
    =  15\nu (\partial_x v_x(x))^2
       \; ,
\end{equation}
which, upon assuming isotropy, equals true dissipation on the
average.

\begin{figure}
\epsfig{file=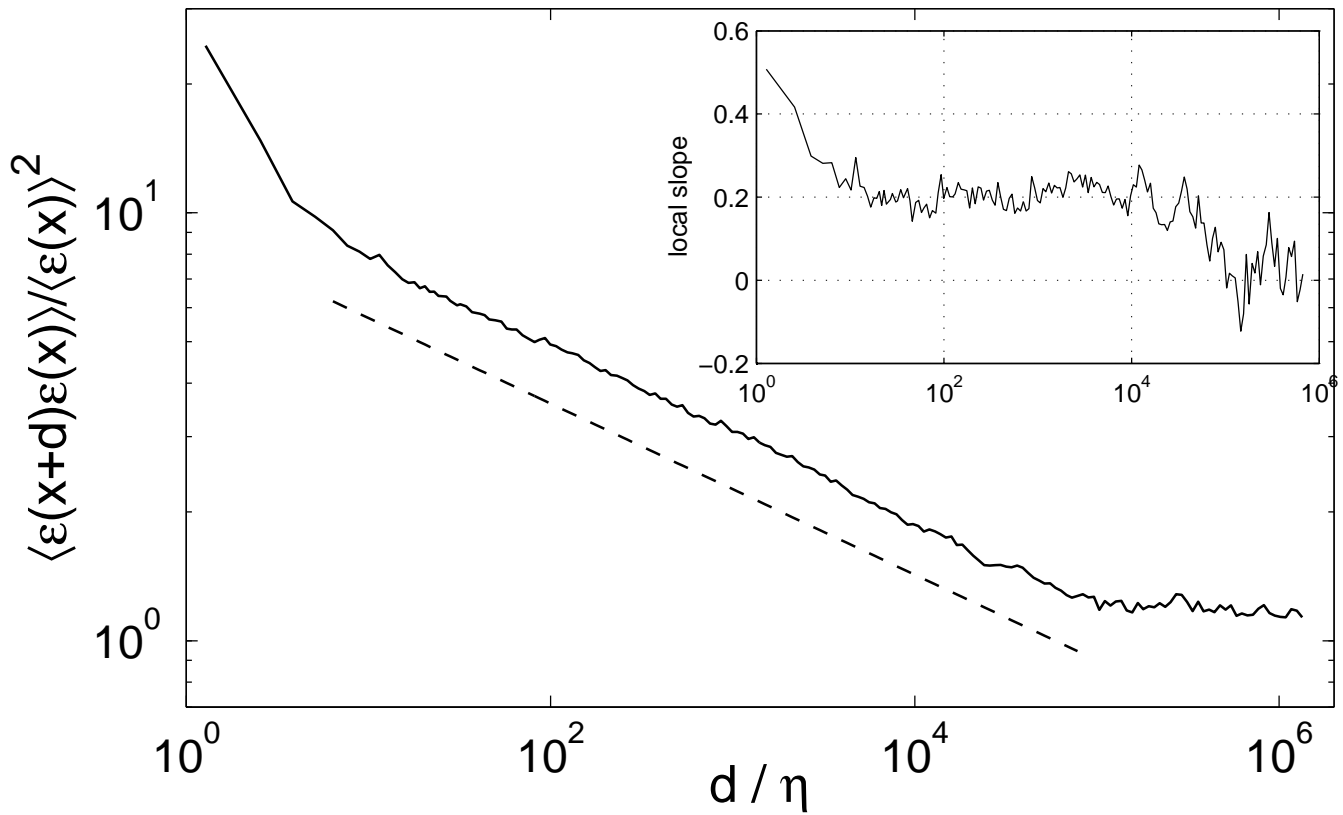,width=\textwidth}
\caption{ Normalized two-point correlation function of the surrogate
energy dissipation field (\ref{two}) obtained via the streamwise
velocity component in the atmosphere. The dashed line has a logarithmic
slope $\mu=0.2$ and its extent indicates the scaling range. Inset:
logarithmic local slope of the two-point correlation function. }
\label{f.one}
\end{figure}

For the extraction of the intermittency exponent, the two-point
correlation function based on the energy dissipation field has
been proposed earlier \cite{MON75}. Figure \ref{f.one} shows the
normalized
two-point correlator $\langle\varepsilon_{\rm
surr1}(x+d)\varepsilon_{\rm surr1}(x)\rangle
/\langle\varepsilon_{\rm surr1}(x)\rangle^2$. 
It reveals a good scaling $\sim d^{-\mu}$ with constant intermittency
exponent, $\mu = 0.20$, over the extended scaling range
$15\eta\leq d \leq 0.3L$, covering most of the
inertial range. This result is in full agreement with older
findings on turbulent jet and atmospheric boundary layer flows
\cite{ANT82,SRE93}, which have also observed a clear and extended
scaling range behaviour for the two-point correlator with about the
same value for the intermittency exponent.

\begin{figure}
\epsfig{file=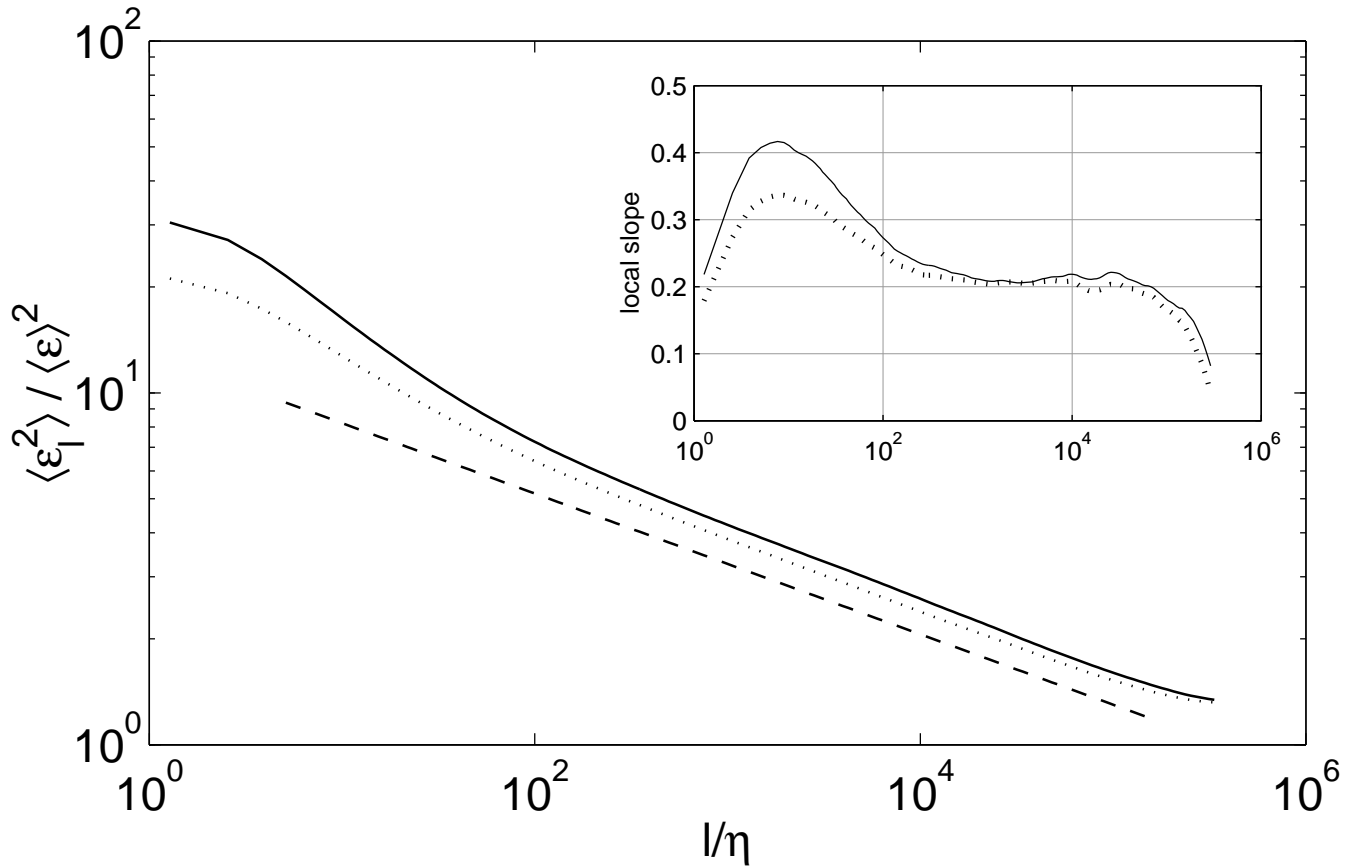,width=\textwidth}
\caption{Second-order integral moment (\ref{three}) based on the
surrogate energy dissipation fields (\ref{two}) (full line), and
(\ref{seven}) (dotted line). The dashed straight line has a logarithmic
slope $\mu=0.2$. Inset shows the logarithmic local slope. }
\label{f.two}
\end{figure}

The two-point correlator of the surrogate energy dissipation
is only one tool for the extraction of the intermittency exponent.
Another approach that is natural from the perspective of the
multifractal picture of turbulence \cite{MAN74,MEN91,FRI95}
employs the second-order moment $\langle\varepsilon_l^2\rangle$ of
the coarse-grained amplitude $\varepsilon_l = l^{-1} \int_l
\varepsilon_{\rm surr}(x) dx$. Figure \ref{f.two} shows $\langle
\varepsilon_l^2 \rangle$ as a function of the averaging scale $l$.
The logarithmic local slope $-d\ln\langle\varepsilon_l^2\rangle /
d\ln{l}$ reveals a scale-independent scaling exponent, which
coincides roughly with the previously found value $\mu=0.20$ for
the intermittency exponent. However, a peculiarity remains: the
logarithmic local slope in Fig. \ref{f.two} is constant (or nearly so) only
in the upper part of the inertial range, and not over the full
range, as found for the two-point correlator. One purpose of
this paper is to understand this puzzle.

To proceed, we note that the second-order moment and the
two-point correlator are closely related (see \cite{MON75}) via
\begin{equation}
\label{three}
  \left\langle \varepsilon_l^2 \right\rangle
    =  \frac{1}{l^2} \int_l dx_1 \int_l dx_2
       \left\langle
       \varepsilon_{\rm surr}(x_1) \varepsilon_{\rm surr}(x_2)
       \right\rangle \;.
\end{equation}
The left-hand-side represents a box integral over the two-point
correlator and may thus be called an integral moment. To see when
the integral moment shows the same scaling exponent as the
two-point correlator, and in what range of scales, we assume the
simplified functional form
\begin{equation}
\label{four}
  \left\langle
  \varepsilon_{\rm surr}(x+d) \varepsilon_{\rm surr}(x)
  \right\rangle
    =  \left\{ \begin{array}{ll}
       c                        &  \;(d<\eta^\prime)                \\
       a (\eta^\prime/d)^\mu    &  \;(\eta^\prime\leq d\leq L^\prime)  \\
       1                        &  \;(d\geq L^\prime)
       \end{array} \right.
\end{equation}
with $\langle\varepsilon_{\rm surr}\rangle=1$, where
$a(\eta^\prime/L^\prime)^\mu=1$ guarantees continuity at the
decorrelation length $d=L^\prime$; the parameter $c$ is left free
for later purposes and $\eta^\prime$ and $L^\prime$ are representative
small and large length scale, respectively.
Upon inserting (\ref{four}) into (\ref{three}) we arrive at
\begin{equation}
\label{five}
  \left\langle \varepsilon_l^2 \right\rangle
    =  \frac{2a}{(1-\mu)(2-\mu)}
       \left( \frac{\eta^\prime}{l} \right)^\mu
       + 2 \left( c - \frac{a}{1-\mu} \right)
       \left( \frac{\eta^\prime}{l} \right)
       + \left( \frac{a}{1-\mu/2} - c \right)
       \left( \frac{\eta^\prime}{l} \right)^2
       \; ,
\end{equation}
valid for $\eta^\prime\leq l\leq L^\prime$. The first term is the
targeted scaling term. The last two terms represent corrections to
rigorous scaling. For increasing $l\gg\eta^\prime$ they fall off
faster than the scaling term. If we were to fine-tune the
leading-order correction to zero, then $c=a/(1-\mu)$. Given that
$\mu=0.20$, the correction to the leading order term also becomes
very small. A closer look at Fig.\ \ref{f.one}, reveals that
a constant $c=a/(1-\mu)$ is too small to approximate the small-scale
behaviour of the two-point correlation.
Consequently, the correction
terms are pronounced for small $l$ and extend far into the
inertial range before becoming negligible. This explains
qualitatively the observed scale-dependence of the second-order
integral moment: only in the upper part of the inertial range does
the scaling term with exponent $\mu$ dominate, whereas for the
lower part strong deviations set in, due to the behaviour of the
two-point correlation function for very small scales.

Figure \ref{f.one} reveals another apparent puzzle: as the two-point
distance approaches $\eta$ the two-point correlation increases
stronger than is suggested by the extrapolation of scaling behaviour.
This is against intuition, since the onset of dissipation is expected
to smooth out the fine structure, instead of building it up. We now
offer a tentative explanation, which reflects the delicate issue of
the surrogacy of the energy dissipation field. The two-point
correlator shown in Fig.\ \ref{f.one} is based on the surrogate energy
dissipation field (\ref{two}) constructed from the longitudinal
velocity component. Other constructions, such as
\begin{eqnarray}
\label{six}
  \varepsilon_{\rm surr2}(x)
    &=&  \frac{15}{2}\nu
         (\partial_x v_y(x))^2
         \; , \\
\label{seven}
  \varepsilon_{\rm surr3}(x)
    &=&  \frac{15}{4}\nu
         \left[
         2 (\partial_x v_x(x))^2
         + (\partial_x v_y(x))^2
         \right]
         \;,
\end{eqnarray}
are also possible. The former is based on the transverse velocity
component alone, whereas the latter combines longitudinal and
transverse components. On average, both constructions are equal in
their mean value to the true energy dissipation field (\ref{one}),
assuming isotropy.

\begin{figure}
\epsfig{file=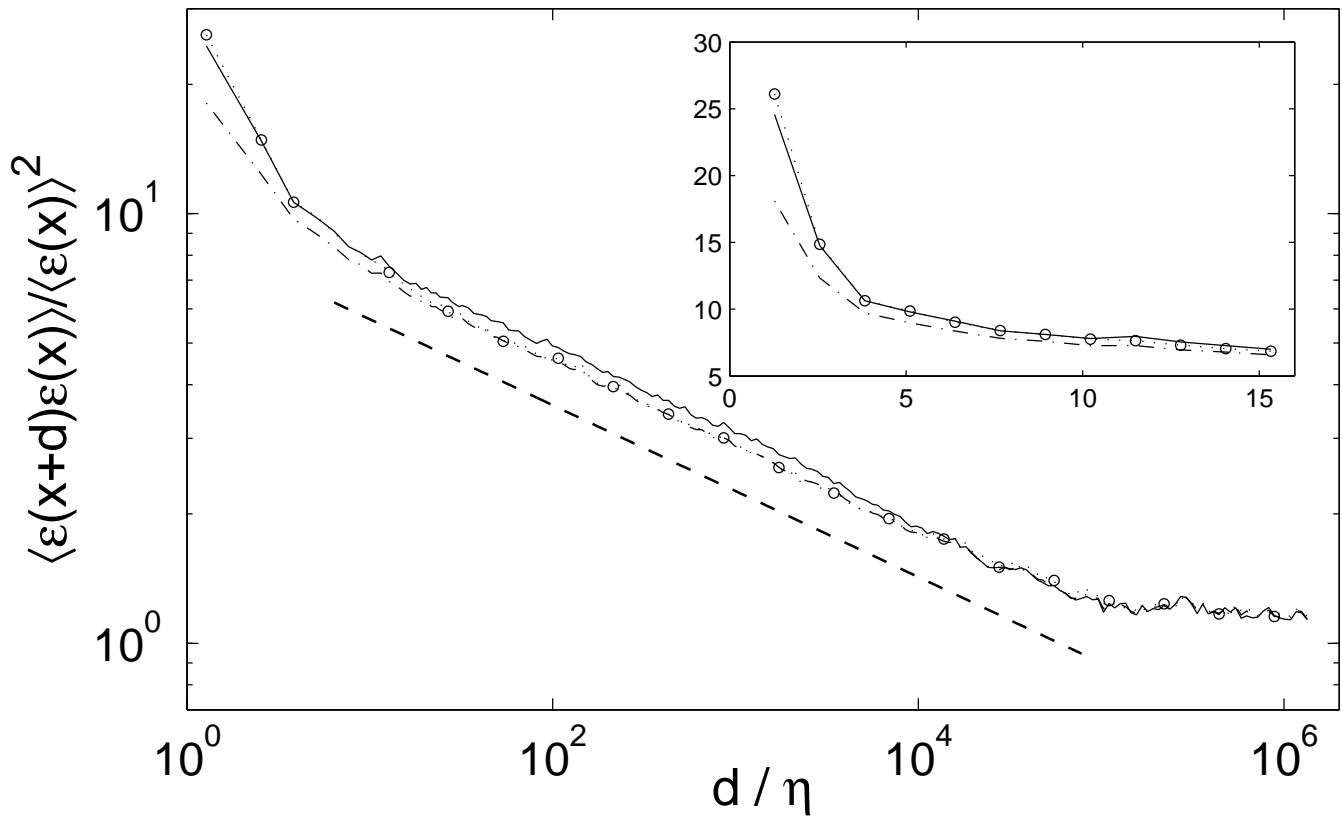,width=\textwidth}
\caption{Normalized
two-point correlation function of the surrogate energy dissipation
fields (\ref{two}) (full line), (\ref{six}) (dotted line with circles), and
(\ref{seven}) (dot-dashed line). The dashed straight line has a
logarithmic slope $\mu=0.2$. Inset magnifies the behaviour for
short separation distances.}
\label{f.three}
\end{figure}

Figure \ref{f.three} compares the two-point correlator obtained from the
surrogate quantities (\ref{two}), (\ref{six}) and (\ref{seven}). 
All three variants reveal a rigorous power-law scaling behaviour within
the extended inertial range $15\eta\leq d\leq 0.3L$ and the
corresponding scaling exponents are within $\mu=0.20\pm 0.01$,
showing little differences. Only for small two-point distances
$d\to\eta$ the two-point correlators begin to differ. Whereas the
variants based on (\ref{two}) and (\ref{six}) practically remain
identical, the two-point correlations based on (\ref{seven}) are
weaker for $d\leq 10\eta$; see inset of Fig.\ \ref{f.three}. In view of the
simplified description (\ref{four}) this implies that the constant
$c$ is smaller for the two-point correlator based on (\ref{seven})
than those based on (\ref{two}) and (\ref{six}). As a consequence
the leading order corrections in the expression (\ref{five}) for
the integral moment also become smaller, so that the scaling term
should begin to dominate even at smaller length scales $l$. Figure
\ref{f.two} confirms this view: the local slope of the integral moment based
on the surrogate field (\ref{seven}) becomes constant at smaller
scales than for (\ref{two}); the upper limit on the scaling range
is the same for the two cases.

When compared to the true energy dissipation (\ref{one}),
the surrogate (\ref{seven}) appears to be closer to
(\ref{one}) than the other two variants (\ref{two}) and
(\ref{six}). We might model the amplitude of the surrogate field in terms
of the amplitude of the true field by the relationship
\begin{equation}
\label{eight}
\varepsilon_{\rm surr}(x)
    = \varepsilon(x) \left( 1+f(x) \right)
      \; ,
\end{equation}
where $f(x)$ behaves as a noise with mean zero. From the defining
equations (\ref{two}), (\ref{six}) and (\ref{seven}) we get
$\varepsilon_{\rm surr3}
 = (\varepsilon_{\rm surr1}+\varepsilon_{\rm surr2})/2
 = \varepsilon [1+(f_1+f_2)/2]$,
so that in comparison with $f_1$ and $f_2$ the noise fluctuation
$f_3=(f_1+f_2)/2$ is reduced. Hence, this allows us to speculate
that if we add more terms from the full list of (\ref{one}),
the extra-strong two-point correlations at small separation
distances $d\leq 15\eta$ reduce further, perhaps even vanish once
the surrogate field has converged to the true field. A quick
numerical investigation, using a shear turbulence code
\cite{SCH01} for $R_\lambda=99$, reveals that the two-point
correlation functions of the surrogate quantities
are identical to that of true energy dissipation, except
for very small distances, where the surrogate field possesses
extra-strong correlations. While a more detailed investigation is
called for, this finding indicates the importance of the subtle
surrogacy issue when interpreting data. It appears clear that the
surrogacy of the energy dissipation field restricts the rigorous
scaling of the second-order integral moment to the upper part of
the inertial range; fortunately, this leaves the rigorous scaling
of the two-point correlation function untouched over that part
of the inertial range.

\begin{figure}
\epsfig{file=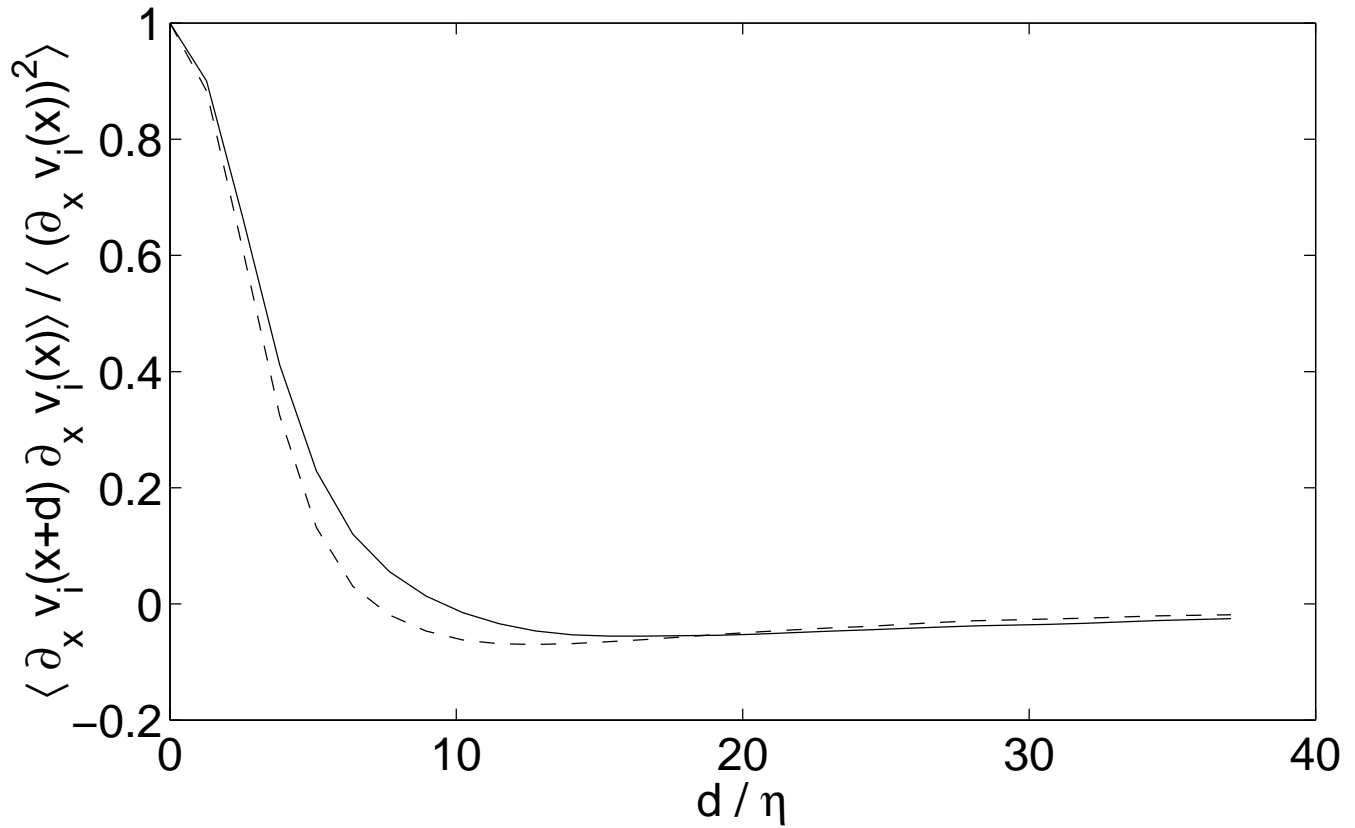,width=\textwidth}
\caption{ Normalized
two-point correlations $\langle (\partial_x v_i(x+d)) (\partial_x
v_i(x)) \rangle / \langle (\partial_x v_i(x))^2 \rangle$ of the
longitudinal (full line, $v_i=v_x$) and transverse (dashed, $v_i=v_y$)
velocity gradients.}
\label{f.four}
\end{figure}

Considering the behaviour of the surrogate correlators for small
separation distances in terms of the modeling relation
(\ref{eight}), the noise field amplitudes cannot be expected to be
uncorrelated, i.e.\ $\langle f(x_1) f(x_2) \rangle \neq \langle
f^2 \rangle \delta(x_1-x_2)$, but should show correlations up to
some separation distance. Empirically, this seems to occur within
the range $|x_1-x_2| \approx 15\eta$. For shorter distances, the
extra-strong correlation sets in for the two-point correlators
based on the surrogate fields. A quantity related to the noise
correlations is the two-point correlation of the velocity gradient
field. Figure \ref{f.four} shows $\langle (\partial_x v_i(x+d))(\partial_x
v_i(x))\rangle$ for the measured longitudinal ($v_i=v_x$) and transverse
($v_i=v_y$) velocity components. All two-point correlations show correlations
up to $d \approx 30\eta$ and become zero for larger distances.

The main message of this simple view on the surrogacy issue of the
energy dissipation field is two-fold: (i) Two-point correlation
functions of the surrogate energy dissipation field reveal
rigorous scaling that is identical to the two-point correlation
function of the true energy dissipation for two-point
distances larger than about fifteen times the dissipation scale.
(ii) The surrogacy of the energy dissipation modifies the
two-point correlations for distances below $d\approx 15\eta$ and
is responsible for restricting the power-law scaling behaviour of
the integral moment to the upper part of the inertial range. A
direct consequence of statements (i) and (ii) is that for the
extraction of the intermittency exponent it is advantageous to use
the two-point correlator. In view of the extended scaling range
$15\eta\leq d\leq 0.3L$ of the two-point correlator of the
energy dissipation, it appears that the intuitive, but
phenomenological, picture of the scale-invariant energy cascade
contains more truth than generally anticipated. Without further
processing such as Extended Self Similarity \cite{BEN93} or
$SO(3)$ decomposition \cite{LVO97,ARA98,KUR00}, conventional
velocity structure functions in shear flows do not show a rigorous
power-law scaling behaviour. This assigns fields such as the energy
dissipation a more fundamental role than that of the velocity
increments. As this inference is based on data analysis alone,
it does not offer any deep theoretical explanation but requires
one.


\acknowledgements
The authors acknowledge fruitful discussions with
J\"urgen Schmiegel, Thomas Dziekan, Hans Eggers, J\"org Schumacher,
Jahanshah Davoudi, Jean-Fran{\c c}ois Pinton and Joachim Peinke.
J.C.\ acknowledges support from DAAD.



\end{document}